\documentclass[twocolumn, prb, notitlepage,floats,superscriptaddress,amsmath,amssymb]{revtex4-2}
\usepackage[version=3]{mhchem} 
\usepackage{bm}
\usepackage[utf8]{inputenc}
\usepackage[T1]{fontenc}
\usepackage{graphicx}
\usepackage{orcidlink}
\usepackage{upgreek}
\usepackage{color}
\usepackage{ulem}
\usepackage{hyperref} 
\hypersetup{colorlinks, citecolor=blue, filecolor=blue, linkcolor=blue, urlcolor=blue}
\usepackage{sidecap}
\usepackage{amssymb}
\usepackage{amsmath}
\usepackage[caption=false]{subfig}
\usepackage{multirow}
\usepackage{setspace}

\def \FUW{University of Warsaw, Faculty of Physics, 02-093 Warsaw, Poland}
\def \PWr{Wrocław University of Science and Technology, Faculty of Fundamental Problems of Technology, 50-730 Wrocław, Poland}
\def \Leipzig {Institute for Inorganic Chemistry, Leipzig University, D-04103 Leipzig, Germany}

\begin{document}
\title{The effect of isotopically-enriched $^{70}$Ge on the Raman scattering in Cr$_2$Ge$_2$Te$_6$}

\author{Grzegorz Krasucki\orcidlink{0000-0000-0000-0000}}
\email{grzegorz.krasucki@fuw.edu.pl}
\affiliation{\FUW}
\author{Katarzyna Olkowska-Pucko\orcidlink{0000-0000-0000-0000}}
\affiliation{\FUW}
\author{Tomasz Wo\'zniak\orcidlink{0000-0000-0000-0000}}
\affiliation{\FUW}
\affiliation{\PWr}
\author{Mihai~I.~Sturza\orcidlink{0000-0000-0000-0000}}
\affiliation{\Leipzig}
\author{Holger Kohlmann}
\affiliation{\Leipzig}
\author{Adam Babi\'nski\orcidlink{0000-0000-0000-0000}}
\email{babinski@fuw.edu.pl}
\affiliation{\FUW}
\author{Maciej R. Molas\orcidlink{0000-0000-0000-0000}}
\email{maciej.molas@fuw.edu.pl}
\affiliation{\FUW}

\begin{abstract}
The highest-energy ${A}_{g}^{5}$ Raman-active vibrational mode in chromium germanium telluride (Cr$_2$Ge$_2$Te$_6$, CGT) exhibits a unique lineshape comprising a series of remarkably narrow peaks. 
While this multiplet was attributed to the natural isotope distribution of germanium affecting \(\text{Ge-Ge}\) pseudo-molecular vibrations, direct evidence has been lacking. 
Here, we present a Raman spectroscopy study of CGT isotopically enriched with \(^{70}\text{Ge}\) to validate this hypothesis. 
Upon isotopic enrichment, both the ${E}_{g}^{5}$ and ${A}_{g}^{5}$ modes exhibit a pronounced blue-shift. 
At room temperature, the ${A}_{g}^{5}$ peak undergoes significant narrowing, while low-temperature measurements reveal a single, well-defined line instead of the previously observed multiplet. 
These observations align precisely with density functional theory calculations, confirming that the multi-line profile originates from Ge isotope distribution. 
Our findings demonstrate that similar isotope-induced Raman splitting may occur in other structurally analogous layered tellurides containing Ge or Si.

\end{abstract}

\maketitle


\section{Introduction}

There has been a growing interest recently in layered magnetic materials, which is driven by their potential applications in spintronics and electronics ~\cite{hoffmann2015, Gibertini2019}. 
Chromium germanium telluride - Cr$_2$Ge$_2$Te$_6$ (CGT), belongs to most prominent layered magnetic compounds.
Its structural stability, compatibility with van der Waals heterostructures, and distinctive electronic and magnetic properties make it a promising candidate for the realization of innovative and advanced device concepts ~\cite{elahi2022}.
CGT is a semiconducting ferromagnet with a bulk Curie temperature of about 61 K, a strong magnetic anisotropy with an out-of-plane easy axis, and negligible coercivity~\cite{Carteaux1995, Ji2013, Zhang2016, Gong2017}.
CGT is also characterized by a strong magneto-elastic coupling, which justifies studies of lattice dynamics in the material with Raman scattering (RS) spectroscopy making it the technique of choice for the nondestructive and versatile analysis of its properties~\cite{Tian2016, Samanta2024, Huang2024, Chakkar2024, Idzuchi2025}.

An intriguing feature of the CGT RS spectrum is the fine structure of the highest-energy ${A}_{g}^{5}$ Raman-active mode.~\cite{Chen, ScRep}
That structure was previously proposed to stem from the natural abundance of germanium, which comprises five stable isotopes: \(^{70}\text{Ge}\), \(^{72}\text{Ge}\), \(^{73}\text{Ge}\), \(^{74}\text{Ge}\), and \(^{76}\text{Ge}\)~\cite{ScRep, NatComm, emergence}. 
Because the ${A}_{g}^{5}$ Raman-active mode involves virtually no involvement of the surrounding lattice atoms~\cite{Milosavljevic}, the vibrations can be modeled as an isolated diatomic Ge pseudo-molecule~\cite{NatComm}.
Fifteen possible isotopic configurations of those pseudo-molecules account for the eleven RS peaks resolved experimentally.~\cite{ScRep, NatComm}
While the excellent agreement with the model strongly supported that assignment, here we definitively validate it by investigating the RS spectra of CGT isotopically enriched with \(^{70}\text{Ge}\). 
For the enriched sample, the fine structure of the ${A}_{g}^{5}$ mode disappears and only one line of the multiplet can be observed at low temperature. 
This peak corresponds precisely to the highest-energy component of the mentioned multiplet, which was previously assigned to the \(^{70}\text{Ge}\)–\(^{70}\text{Ge}\) pseudomolecule configuration. 
Additionally, the ${E}_{g}^{5}$  mode exhibits a pronounced blue-shift as a result of the \(^{70}\text{Ge}\) enrichment . 
These experimental findings are fully consistent with density functional theory (DFT) calculations.\newline

\section{Results}

\begin{figure}[t]
		\subfloat{}%
		\centering
		\includegraphics[width=1\linewidth]{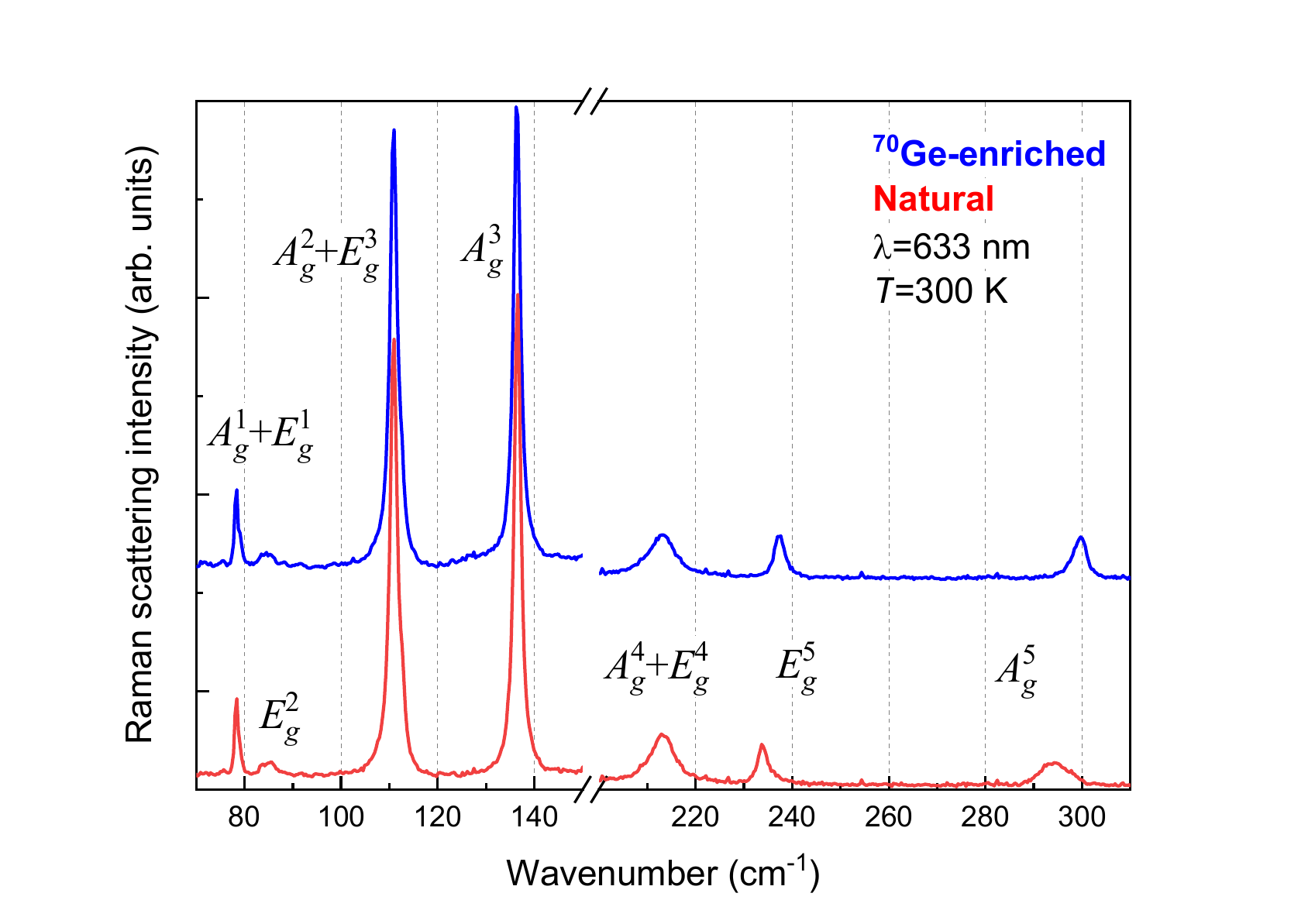}
        \caption{Room temperature Raman scattering spectra of CGT excited with $\lambda$=633 nm laser. The spectra of natural and $^{70}$Ge-enriched CGT are vertically shifted for more clarity. 
        Dotted lines are included to emphasize the effect of $^{70}$Ge enrichment on the modes.}
		\label{fig:2}
\end{figure}

The room-temperature RS spectra of both: the natural and the $^{70}$Ge-enriched CGT are presented in Fig.~\ref{fig:2}.

No apparent effect of the $^{70}$Ge-enrichment on the RS line-shape can be observed in the spectral range below 220~cm$^{-1}$.
The observed features correspond to the previously identified Raman-ctive modes~\cite{ScRep}. 
These were: 
${A}_{g}^{1}$ and ${E}_{g}^{1}$ at 78.3~cm$^{-1}$ and 79.1~cm$^{-1}$, 
${E}_{g}^{2}$ at 84.5~cm$^{-1}$, 
${A}_{g}^{2}$+${E}_{g}^{3}$ at 110.9~cm$^{-1}$ and 112.6~cm$^{-1}$, 
${A}_{g}^{3}$ at 136.4~cm$^{-1}$, and
${A}_{g}^{4}$+${E}_{g}^{4}$ at 213~cm$^{-1}$.
The line-shape of the spectrum of the $^{70}$Ge-enriched sample is characteristic for CGT and confirms its high quality.

The effect of $^{70}$Ge enrichment on the RS spectrum can be clearly observed in the RS spectrum at higher energies.
The ${E}_{g}^{5}$ and ${A}_{g}^{5}$
modes observed in the natural CGT at 233~cm$^{-1}$ and 294.4~cm$^{-1}$ respectively, blueshift in the $^{70}$Ge-enriched sample to peak at 
237.4~cm$^{-1}$ and 300.1~cm$^{-1}$, respectively.
Moreover, a substantial narrowing of the ${A}_{g}^{5}$ mode can be observed from 7~cm$^{-1}$ in the natural crystal to 3.4~cm$^{-1}$ in the $^{70}$Ge-enriched one.

\begin{figure}[t]
		\subfloat{}%
		\centering
		\includegraphics[width=1 \linewidth]{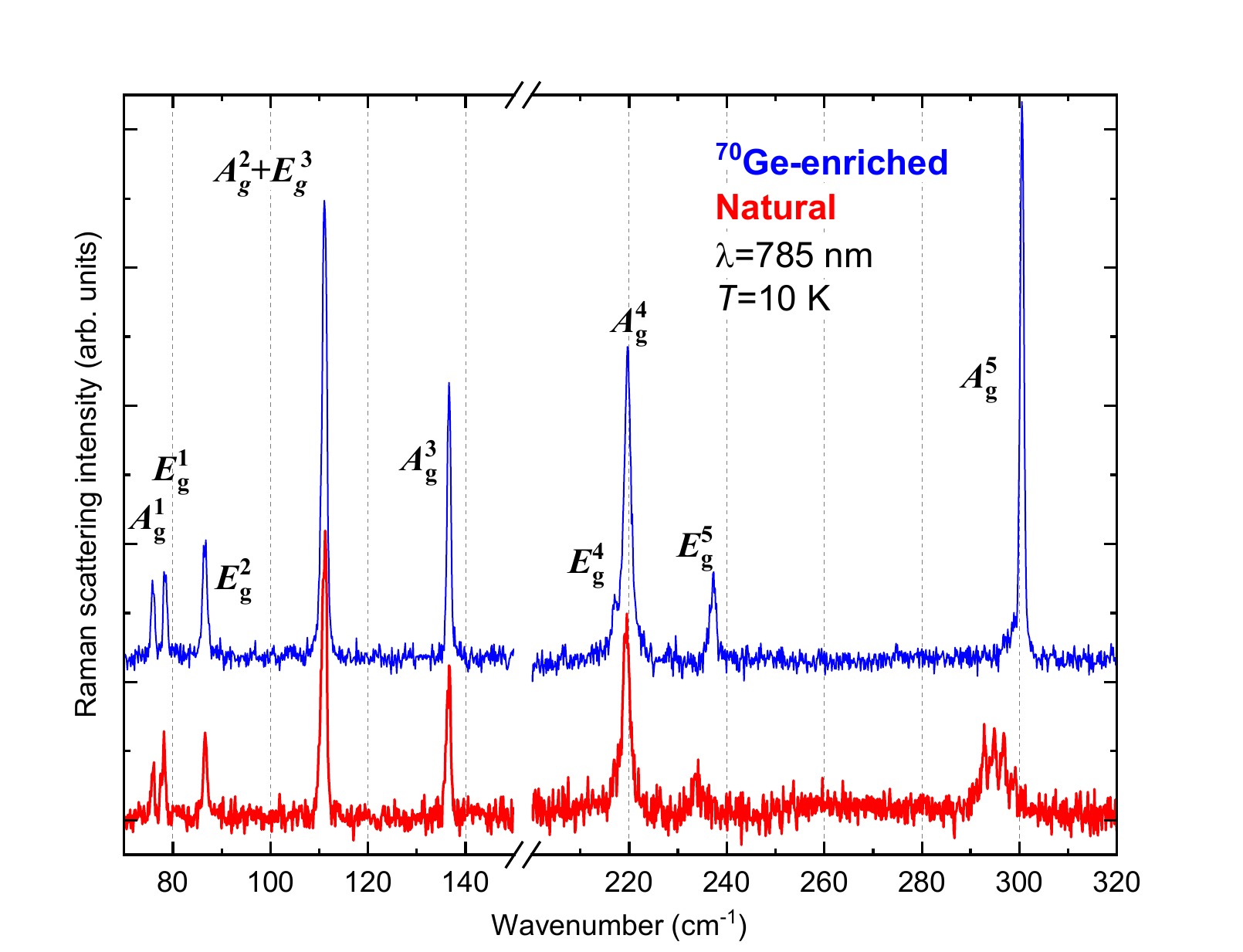}
        \caption{Low-temperature ($T$=10 K) Raman scattering spectra of CGT excited with $\lambda$=785 nm laser. 
        The spectra from natural and $^{70}$Ge-enriched CGT are vertically shifted for more clarity.}
		\label{fig:3}
\end{figure}

The effects are more apparent in the low-temperature ($T$=10 K) RS spectra measured with $\lambda$=785 nm - see Fig.~\ref{fig:3}. 
No significant difference can be noticed between the lineshape of the spectra below 150~cm$^{-1}$ from the natural and $^{70}$Ge-enriched samples except for for changes in their relative intensities.
The peaks observed at low temperature have smaller linewidth than those measured at room temperature.
This allows in particular to clearly distinguish the well-separated ${A}_{g}^{1}$ and ${E}_{g}^{1}$ peaks in both samples at 75.9~cm$^{-1}$ and 78.5~cm$^{-1}$ respectively. 
Moreover the ${E}_{g}^{2}$ peak can be clearly observed at 86.6~cm$^{-1}$.

\begin{figure}[!th]
		\subfloat{}%
		\centering
		\includegraphics[width=1\linewidth]{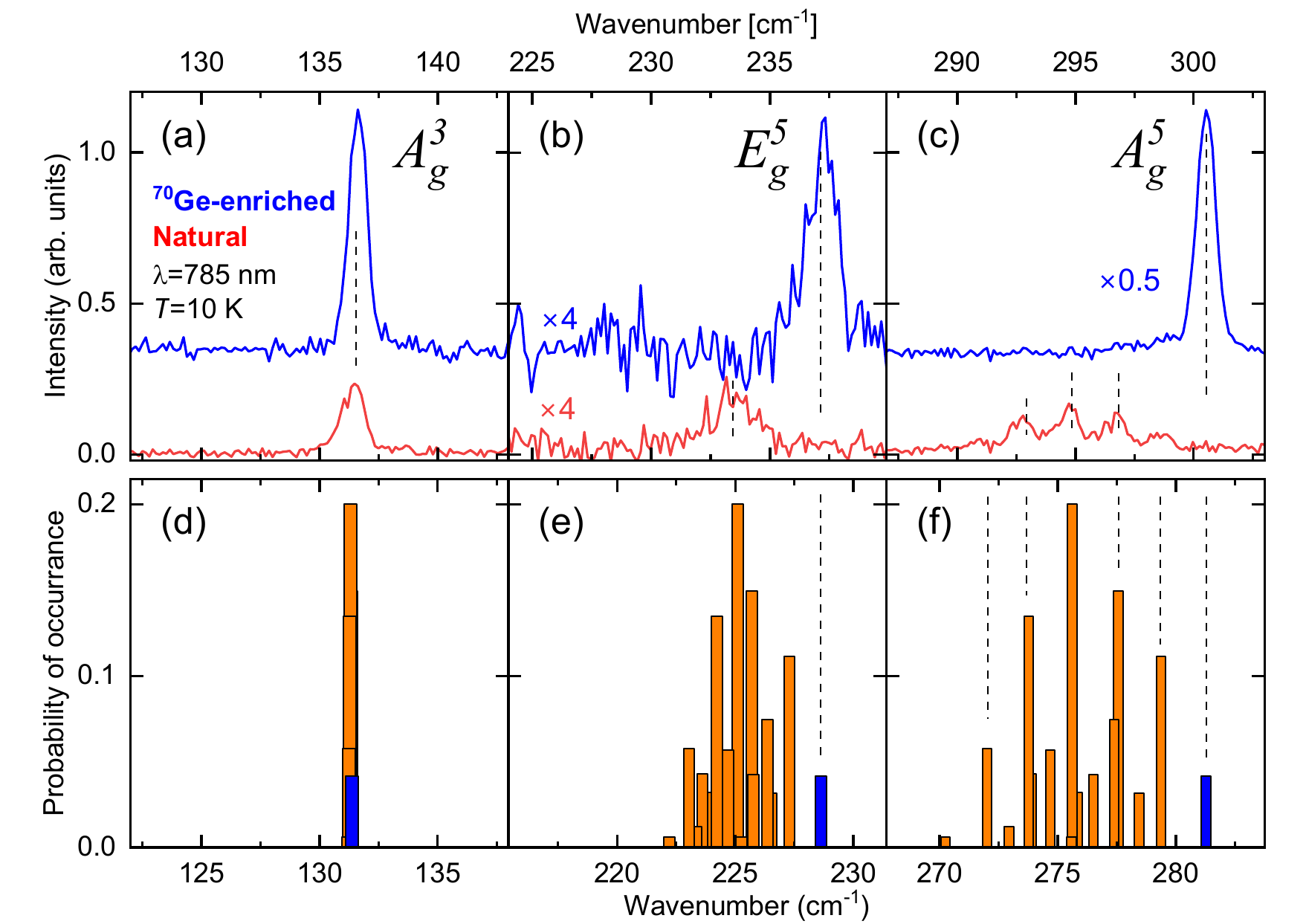}
        \caption{Details of the low temperature ($T$=10 K) Raman scattering spectra. The ${A}_{g}^{3}$ (a), ${E}_{g}^{5}$ (b) and ${A}_{g}^{5}$ (c) modes in the natural and $^{70}$Ge-enriched CGT. 
        The probability of occurrence for particular Ge-Ge pseudo-molecules giving rise to vibrations of particular energy due to the ${A}_{g}^{3}$ (d), ${E}_{g}^{5}$ (e) and ${A}_{g}^{5}$ (f) modes in the natural CGT crystal. The probability for $^{70}$Ge-$^{70}$Ge pseudo-molecule is shown in blue.}
		\label{fig:4}
\end{figure}

There were also
${A}_{g}^{2}$+${E}_{g}^{3}$ 
features de-convoluted to peak at 111.0~cm$^{-1}$ and 111.4~cm$^{-1}$ 
and ${A}_{g}^{3}$ at 136.6~cm$^{-1}$ modes observed in the spectra.
All those modes are not clearly affected by the $^{70}$Ge-enrichment.
On the contrary, only in the spectrum from the $^{70}$Ge-enriched sample, the ${E}_{g}^{4}$ and ${A}_{g}^{4}$ modes can be distinguished at 217~cm$^{-1}$ and 219.7~cm$^{-1}$ respectively.
As observed in room-temperature measurements, the ${E}_{g}^{5}$ mode shifts from 233.7~cm$^{-1}$ in the natural CGT to 237.3~cm$^{-1}$ in the $^{70}$Ge enriched sample.
The most spectacular is the substantial narrowing of the ${A}_{g}^{5}$ and its blueshift.
Details of the ${A}_{g}^{5}$ lineshape in natural and $^{70}$Ge-enriched sample can be appreciated in Fig.~\ref{fig:4}c.  
It can be clearly observed that the ${A}_{g}^{5}$ peak in the $^{70}$Ge-enriched CGT corresponds to the highest energy range of the multiline structure of the ${A}_{g}^{5}$ peak in natural crystal.
In our previous contribution~\cite{NatComm}, the multiplet was attributed to the isotope variety of Ge atoms present in natural CGT (for the review on the effect of isotopes on the properties of semiconductors - see~\cite{Cardona}).
We used the abundance of the isotopes ~\cite{deLaeter2003} to model the multiline spectrum of the ${A}_{g}^{5}$ mode within the spring-and-ball model showing excellent agreement with experimental data in terms of both expected energy distribution and line intensities.
The highest-energy component of the multiplet was attributed to normal vibrations of the $^{70}$Ge-$^{70}$Ge pseudo-molecule.

In order to account for our experimental observation in more formal way, we performed DFT calculations to find the energies of particular CGT normal vibration for a specific pair of Ge isotopes giving rise to the modes.
The results of calculations for three modes: (${A}_{g}^{3}$, ${E}_{g}^{5}$, and ${A}_{g}^{5}$) are presented in Table~\ref{Tbs:Theory}.

\begin{table}[]
\caption
    {The energies of the ${A}_{g}^{3}$, ${E}_{g}^{5}$, and ${A}_{g}^{5}$ modes in CGT calculated using DFT.
     Energies of modes involving particular Ge isotopes ($^{70}$Ge, $^{72}$Ge, $^{73}$Ge, $^{74}$Ge, and $^{76}$Ge) building the Ge pseudo-molecule. Weighted mean energies are also shown.}
\centering
\begin{tabular}[t]{c|c|c|c}
\hline
      &${A}_{g}^{3}$ & ${E}_{g}^{5}$  & ${A}_{g}^{5}$\\
\hline
    & \multicolumn{3} {c} {Wavenumber (cm$^{-1}$)} \\
\hline \hline
      $^{70}$Ge -$^{70}$Ge &131.379 & 228.634& 281.268\\
      $^{70}$Ge -$^{72}$Ge &131.354 & 227.294& 279.363\\
      $^{70}$Ge -$^{73}$Ge &131.341 & 226.518& 278.446\\
      $^{70}$Ge -$^{74}$Ge &131.329 & 225.701& 277.556\\
      $^{70}$Ge -$^{76}$Ge &131.302 & 223.996& 275.84 \\
      $^{72}$Ge -$^{72}$Ge &131.33  & 226.362& 277.38 \\
      $^{72}$Ge -$^{73}$Ge &131.317 & 225.768& 276.511\\
      $^{72}$Ge -$^{74}$Ge &131.305 & 225.103& 275.611\\
      $^{72}$Ge -$^{76}$Ge &131.279 & 223.615& 273.876\\
      $^{73}$Ge -$^{73}$Ge &131.305 & 225.272& 275.579\\
      $^{73}$Ge -$^{74}$Ge &131.292 & 224.699& 274.674\\
      $^{73}$Ge -$^{76}$Ge &131.267 & 223.356& 272.93 \\
      $^{74}$Ge -$^{74}$Ge &131.28  & 224.219& 273.765\\
      $^{74}$Ge -$^{76}$Ge &131.254 & 223.041& 272.011\\
      $^{76}$Ge -$^{76}$Ge &131.229 & 222.21 & 270.239\\
\hline \hline
         Weighted mean &131.306 & 225.153  & 275.789\\
\hline
 	 	
      \end{tabular}

\label{Tbs:Theory}
\end{table}

The energies and the probability of occurrence for corresponding Ge pairs, which corresponds to the line intensity are also presented in separate panels (d,e,f) of Fig.~\ref{fig:4}.
One may note that although the absolute values of phonon energies theoretically established within DFT are shifted from the experimental ones~\cite{DFTAccuracy}, the effect of atomic mass on the mode distribution properly reflects the experimental data.
As can be seen in Fig.~\ref{fig:4}f, the components of ${A}_{g}^{5}$ in natural CGT are well energy-separated, which allows for their differentiation in the spectrum.
When the theoretically predicted peak due to $^{70}$Ge-$^{70}$Ge vibrations is set to correspond to the experimentally observed peak in the $^{70}$Ge-enriched sample, the correspondence also holds for other components of the ${A}_{g}^{5}$ multiplet.
This confirms that our model of isotope-related ${A}_{g}^{5}$ multiplet structure is correct as just one line due to the $^{70}$Ge-$^{70}$Ge pseudo-molecule can be observed in the $^{70}$Ge-enriched sample instead of the nultiplet characteristic for the natural CGT.
The observed effect of $^{70}$Ge enrichment on the ${E}_{g}^{5}$ mode is also supported by our DFT calculations. 
The isotope-related energy distribution for the mode is smaller, which prevents the identification of isotope-specific ${E}_{g}^{5}$ components.
However. its experimentally observed blue-shift is well-reproduced by our DFT calculations (compare Fig.~\ref{fig:4}b and e).
The calculations also point out to a negligible effect of the $^{70}$Ge enrichment on the energies of other RS modes.
As an example, the spectrum of ${A}_{g}^{3}$ in natural and $^{70}$Ge-enriched CGT can be compared (see Fig.~\ref{fig:4}a) with the energy distribution due to isotope composition of Ge-Ge pairs (see Fig.~\ref{fig:4}d).
Theoretically predicted, isotope-specific broadening of the peak is substantially lower than that for previously analyzed modes.
Similar effect was noticed for all the Raman-active modes of lower energy.
The mass of Ge atom does not affect noticeably their energies. 

\section{Conclusions}

We have investigated the RS of $^{70}$Ge-enriched and natural CGT at room temperature and at $T$=10K.
We have shown that the highest-energy ${A}_{g}^{5}$ Raman-active mode in the $^{70}$Ge-enriched CGT has no fine structure.
A single peak due to the ${A}_{g}^{5}$ mode has been observed, which corresponds to the highest-energy component of the ${A}_{g}^{5}$ multiplet previously attributed to vibrations of the $^{70}$Ge-$^{70}$Ge pseudo-molecule.
This confirms the validity of our previous attribution of the ${A}_{g}^{5}$ mode fine structure to isotopic diversity of natural germanium.
We have also shown a significant blueshift of the ${E}_{g}^{5}$ mode as a result of the $^{70}$Ge enrichment, which agrees with the results of DFT calculations.
In accordance with the calculations, no other Raman-active mode was noticeably affected by the enrichment.
We believe that our findings will draw renewed attention to isotopically-enriched vdW materials. 
The CGT crystal structure is shared by numerous materials with diverse properties. 
While the fine structure of the Raman-active mode corresponding to Si pseudo-molecule vibrations-reflecting the isotope composition of natural silicon - has already been observed in CrSiTe$_3$~\cite{Chen} and InSiTe$_3$~\cite{Belojica}, many related materials, including InGeTe$_3$~\cite{Zhao}, BiSiTe$_3$ ~\cite{Luo} or SbSiTe$_3$~\cite{LuoSb}, remain to be investigated to reveal the possible effects of Ge/Si isotopes on their fundamental properties.\newline
 
\section{Methods}

\subsection{Sample growth}

A self-flux method, leveraging a germanium-tellurium rich flux, based on established procedures~\cite{Zeisner2019} was employed to synthesize single CGT crystals using isotopically-enriched $^{70}$Ge or normally occurring Ge for a reference sample.

\subsection{Optical spectrosocpy}

The RS spectra of two CGT samples were investigated: 
(a) the sample grown with naturally occurring Ge (for more details see Ref.\cite{ScRep}), which will be referred in the following to as "natural" CGT and 
(b) the sample grown with isotopically enriched $^{70}$Ge, which will be referred to as "$^{70}$Ge-enriched" CGT.
Both samples were a few hundreds microns thick, which allows to treat them as bulk-like CGT flakes.
The CGT crystals were placed on a Si substrate. 
Room-temperature RS measurements were conducted in HR LabRAM Evolution system with $\lambda =$ 633 nm laser excitation. 
Measurements were performed with excitation light focused using a 100× long-working-distance objective.
Low temperature RS experiments were conducted using $\lambda=$ 785 nm laser excitation on the sample placed on a cold finger of a continuous flow cryostat mounted on x–y motorized positioners.
The excitation light was focused using a 50× long-working-distance objective with a numerical aperture (NA) of 0.55, producing a spot of approximately diameter 1 $\mu$m. 
The signal was collected in back-scattering geometry through the same microscope objective and detected using a liquid-nitrogen-cooled charge-coupled device (CCD) camera. 
The excitation power focused on the samples in both cases was kept at approx. 1~mW or less in all studies.

\subsection{Theory} 

DFT calculations were performed in Vienna ab-initio simulation package (VASP) version 6.4.2~\cite{KRESSE199615, PhysRevB.54.11169}. 
The projector augment wave (PAW) potentials Cr\_pv, Ge\_d and Te with Pedew-Burke-Ernzerhof (PBE) parametrization of the general gradients approximation (GGA) to the exchange-correlation functional were used \cite{ PhysRevB.59.1758}. 
The unit cell parameters and atomic positions were optimized with criteria of 10$^{-5}$ ~eV/Å for forces and 0.1 kbar for stresses, including the D3 correction to van der Waals interactions \cite{10.1063/1.3382344}. 
An energy cut-off of 400 eV and a 6$\times$6$\times$6 $\Gamma$-centered k-points grid were used. 
Phonon calculations were performed within the direct method as implemented in Phonopy \cite{PhysRevLett.78.4063,phonopy-phono3py-JPSJ,phonopy-phono3py-JPCM}. 
A 2$\times$2$\times$2 supercell was used to evaluate the interatomic force constants.

\subsection*{Acknowledgements}
The work has been supported by the National Science Centre, Poland (Grants Nos 2020/37/B/ST3/02311, 2025/57/B/ST5/03288). The Authors are grateful to Prof. A.Korgul and Prof. M.Karny of the Faculty of Physics, University of Warsaw, for fruitful discussions and support of this work. DFT calculations have been carried out using computational resources of the Interdisciplinary Centre for Mathematical and Computational Modelling (ICM), University of Warsaw, under grant GB83-21.

\subsection*{ Author contributions}
A.B and M. R. M. conceived and managed the project. 
The CGT samples were grown by M.I.S. and H.K..
Raman scattering spectroscopy was performed by G.K., K.O.-P.. and M. R. M..
G.K., M.R.M., T.W., and A.B. carried out the data analysis. 
DFT calculations were done by T.W..
The paper was written by A.B. with input from all coauthors.

\subsection*{ Competing interests}
The authors declare no competing interests.

\subsection*{Data availability}
The data that support the findings of this study are openly available in [repository name] at ..., reference number [reference number].

\bibliographystyle{apsrev4-2}
\bibliography{biblio.bib}

\end{document}